\documentclass[12pt,a4paper]{article}
\makeatletter
\def\eqnarray{%
\stepcounter{equation}%
\let\@currentlabel=\theequation
\global\@eqnswtrue
\global\@eqcnt\z@
\tabskip\@centering
\let\\=\@eqncr
$$\halign to \displaywidth\bgroup\@eqnsel\hskip\@centering
$\displaystyle\tabskip\z@{##}$&\global\@eqcnt\@ne
\hfil$\displaystyle{{}##{}}$\hfil
&\global\@eqcnt\tw@$\displaystyle\tabskip\z@{##}$\hfil
\tabskip\@centering&\llap{##}\tabskip\z@\cr}
\makeatother
\newcommand{\ket}[1]{{\vert{#1}\rangle}}
\newcommand{\bra}[1]{{\langle{#1}\vert}}
\newcommand{\braket}[1]{{\langle{#1}\vert{#1}\rangle}}

\newcommand{\fukuso}{{\mathbf C}}

\begin{document}

\title{\sl Comment on ``A Recursive Parametrisation of Unitary Matrices"}
\author{
  Kazuyuki FUJII\\
  \thanks{E-mail address : fujii@yokohama-cu.ac.jp }
  Department of Mathematical Sciences\\
  Yokohama City University\\
  Yokohama, 236--0027\\
  Japan
  }
\date{}
\maketitle
%
%
%
%
\begin{abstract}
  In this note we make a comment on the paper by Jarlskog (math--ph/0504049) 
  in which she gave a parametrisation to unitary matrices. Namely, we show 
  that her recursive method is essentially obtained by the canonical 
  coordinate of the second kind in the Lie group theory, and propose 
  a problem on constructing unitary gates in quantum computation by making 
  use of the parametrisation.
\end{abstract}
%


%
%
%
%

In the paper \cite{CJ} Jarlskog gave an interesting parametrisation to 
unitary matrices. See also \cite{Dita} as a similar parametrisation. 
In this note we show that her recursive method is essentially obtained by the 
so--called canonical coordinate of the second kind in the Lie group theory. 

The unitary group is defined as 
\begin{equation}
U(n)=\left\{U \in M(n,\fukuso)\ |\ U^{\dagger}U=UU^{\dagger}={\bf 1}_{n}
\right\}
\end{equation}
and its (unitary) algebra is given by
\begin{equation}
u(n)=\left\{X \in M(n,\fukuso)\ |\ X^{\dagger}=-X \right\}.
\end{equation}
Then the exponential map is
\begin{equation}
\mbox{exp} : u(n)\ \longrightarrow\ U(n)\ ;\ X\ \mapsto\ U\equiv \mbox{e}^{X}.
\end{equation}
This map is canonical but not easy to calculate.

We write down the element $X \in u(n)$ explicitly :
\begin{equation}
X=
\left(
\begin{array}{cccccc}
i\theta_{1} & z_{12} & z_{13} & \cdots & z_{1,n-1} & z_{1n} \\
-\bar{z}_{12} & i\theta_{2} & z_{23} & \cdots & z_{2,n-1} & z_{2n} \\
-\bar{z}_{13} & -\bar{z}_{23} & i\theta_{3} & \cdots & z_{3,n-1} & z_{3n} \\
 \vdots & \vdots & \vdots & \ddots & \vdots & \vdots  \\
-\bar{z}_{1,n-1} & -\bar{z}_{2,n-1} & -\bar{z}_{3,n-1} & \cdots & 
i\theta_{n-1} & z_{n-1,n} \\
-\bar{z}_{1,n} & -\bar{z}_{2,n} & -\bar{z}_{3,n} & \cdots & -\bar{z}_{n-1,n}
& i\theta_{n}
\end{array}
\right).        
\end{equation}
This $X$ is decomposed into
\[
X=X_{0}+X_{2}+\cdots + X_{j}+\cdots +X_{n}
\]
where
\begin{equation}
X_{0}=
\left(
\begin{array}{cccccc}
i\theta_{1} & & & & &   \\
& i\theta_{2} & & & &   \\
& & i\theta_{3} & & &   \\
& & & \ddots &  &       \\
& & & & i\theta_{n-1} & \\
& & & & & i\theta_{n} 
\end{array}
\right)
\end{equation}
and for $2\leq j\leq n$
\begin{equation}
X_{j}=
\left(
\begin{array}{cccccc}
0 & & & & &   \\
& \ddots & & \ket{z_{j}} & &   \\
& & \ddots & & &   \\
& -\bra{z_{j}} &  & 0 &  &  \\
& & & & \ddots & \\
& & & & & 0
\end{array}
\right),\quad 
\ket{z_{j}}=
\left(
\begin{array}{c}
z_{1j} \\
z_{2j} \\
\vdots \\
z_{j-1,j}
\end{array}
\right).
\end{equation}

Then the canonical coordinate of the second kind in the unitary group (Lie 
group) is well--known and given by
\begin{equation}
u(n) \ni X=X_{0}+X_{2}+\cdots + X_{j}+\cdots +X_{n}
\ \longrightarrow\ 
\mbox{e}^{X_{0}}\mbox{e}^{X_{2}}\cdots \mbox{e}^{X_{j}}\cdots \mbox{e}^{X_{n}}
 \in U(n)
\end{equation}
in this case \footnote{There are of course some variations}. Therefore 
let us calculate $\mbox{e}^{X_{j}}$ for $j\geq 2$ ($j=0$ is trivial). From
\[
X_{j}=
\left(
\begin{array}{ccc}
{\bf 0}_{j-1} & \ket{z_{j}} &   \\
-\bra{z_{j}} & 0 &              \\
& & {\bf 0}_{n-j}
\end{array}
\right)
\equiv 
\left(
\begin{array}{cc}
K &               \\
  & {\bf 0}_{n-j}
\end{array}
\right)
\quad \Longrightarrow \quad
\mbox{e}^{X_{j}}=
\left(
\begin{array}{cc}
 \mbox{e}^{K} &     \\
   & {\bf 1}_{n-j}
\end{array}
\right)
\]
we have only to calculate the term $\mbox{e}^{K}$, which is easy task. 
From
\begin{eqnarray}
K&=&
\left(
\begin{array}{cc}
{\bf 0}_{j-1} & \ket{z_{j}} \\
-\bra{z_{j}}  & 0
\end{array}
\right),\quad 
K^{2}=
\left(
\begin{array}{cc}
-\ket{z_{j}}\bra{z_{j}} &  \\
  & -\braket{z_{j}}
\end{array}
\right),   \nonumber \\
K^{3}&=&
\left(
\begin{array}{cc}
{\bf 0}_{j-1} & -\braket{z_{j}}\ket{z_{j}}  \\
\braket{z_{j}}\bra{z_{j}} & 0            
\end{array}
\right)
=-\braket{z_{j}}K
\end{eqnarray}
we have important relations
\[
K^{2n+1}=\left(-\braket{z_{j}}\right)^{n}K,\quad 
K^{2n+2}=\left(-\braket{z_{j}}\right)^{n}K^{2}
\quad \mbox{for}\quad n\geq 0.
\]
Therefore
\begin{eqnarray*}
\mbox{e}^{K}
&=&{\bf 1}_{j}+\sum_{n=0}^{\infty}\frac{1}{(2n+2)!}X^{2n+2}
              +\sum_{n=0}^{\infty}\frac{1}{(2n+1)!}X^{2n+1}  \\
&=&{\bf 1}_{j}
+\sum_{n=0}^{\infty}\frac{1}{(2n+2)!}\left(-\braket{z_{j}}\right)^{n}K^{2}
+\sum_{n=0}^{\infty}\frac{1}{(2n+1)!}\left(-\braket{z_{j}}\right)^{n}K  \\
&=&{\bf 1}_{j}
+\sum_{n=0}^{\infty}(-1)^{n}\frac{\left(\sqrt{\braket{z_{j}}}\right)^{2n}}
{(2n+2)!}K^{2}
+\sum_{n=0}^{\infty}(-1)^{n}\frac{\left(\sqrt{\braket{z_{j}}}\right)^{2n}}
{(2n+1)!}K  \\
&=&{\bf 1}_{j}
-\frac{1}{\braket{z_{j}}}\sum_{n=0}^{\infty}(-1)^{n+1}
\frac{\left(\sqrt{\braket{z_{j}}}\right)^{2n+2}}{(2n+2)!}K^{2}
+\frac{1}{\sqrt{\braket{z_{j}}}}\sum_{n=0}^{\infty}(-1)^{n}
\frac{\left(\sqrt{\braket{z_{j}}}\right)^{2n+1}}{(2n+1)!}K  \\
&=&{\bf 1}_{j}
-\frac{1}{\braket{z_{j}}}\sum_{n=1}^{\infty}(-1)^{n}
\frac{\left(\sqrt{\braket{z_{j}}}\right)^{2n}}{(2n)!}K^{2}
+\frac{1}{\sqrt{\braket{z_{j}}}}\sum_{n=0}^{\infty}(-1)^{n}
\frac{\left(\sqrt{\braket{z_{j}}}\right)^{2n+1}}{(2n+1)!}K  \\
&=&{\bf 1}_{j}
-\frac{1}{\braket{z_{j}}}\left(\mbox{cos}(\sqrt{\braket{z_{j}}})-1\right)K^{2}
+\frac{\mbox{sin}(\sqrt{\braket{z_{j}}})}{\sqrt{\braket{z_{j}}}}K  \\
&=&{\bf 1}_{j}
+\left(1-\mbox{cos}(\sqrt{\braket{z_{j}}})\right)\frac{1}{\braket{z_{j}}}K^{2}
+\mbox{sin}(\sqrt{\braket{z_{j}}})\frac{1}{\sqrt{\braket{z_{j}}}}K.
\end{eqnarray*}

If we define a normalized vector as
\[
\ket{\tilde{z}_{j}}=\frac{1}{\sqrt{\braket{z_{j}}}}\ket{z_{j}}
\ \Longrightarrow\ \braket{\tilde{z}_{j}}=1
\]
then 
\[
\frac{1}{\sqrt{\braket{z_{j}}}}K
=
\left(
\begin{array}{cc}
{\bf 0}_{j-1} & \ket{\tilde{z}_{j}}   \\
-\bra{\tilde{z}_{j}} & 0 
\end{array}
\right),\quad 
\frac{1}{\braket{z_{j}}}K^{2}
=
\left(
\begin{array}{cc}
-\ket{\tilde{z}_{j}}\bra{\tilde{z}_{j}} &  \\
  & -1
\end{array}
\right).
\]
Therefore
\begin{equation}
\mbox{e}^{K}
=
\left(
\begin{array}{cc}
{\bf 1}_{j-1}-\left(1-\mbox{cos}(\sqrt{\braket{z_{j}}})\right)
\ket{\tilde{z}_{j}}\bra{\tilde{z}_{j}}& 
\mbox{sin}(\sqrt{\braket{z_{j}}})\ket{\tilde{z}_{j}}   \\
-\mbox{sin}(\sqrt{\braket{z_{j}}})\bra{\tilde{z}_{j}} & 
\mbox{cos}(\sqrt{\braket{z_{j}}})
\end{array}
\right).
\end{equation}

As a result we obtain
\begin{equation}
\label{eq:fundamental}
\mbox{e}^{X_{j}}
=
\left(
\begin{array}{ccc}
{\bf 1}_{j-1}-\left(1-\mbox{cos}(\sqrt{\braket{z_{j}}})\right)
\ket{\tilde{z}_{j}}\bra{\tilde{z}_{j}}& 
\mbox{sin}(\sqrt{\braket{z_{j}}})\ket{\tilde{z}_{j}} &  \\
-\mbox{sin}(\sqrt{\braket{z_{j}}})\bra{\tilde{z}_{j}} & 
\mbox{cos}(\sqrt{\braket{z_{j}}}) & \\
  &  &  {\bf 1}_{n-j}
\end{array}
\right).
\end{equation}
This is just the matrix given in \cite{CJ}.

\par \vspace{3mm}
We make a comment on the case of $n=2$. Since 
\[
\ket{\tilde{z}}=\frac{z}{|z|}\equiv \mbox{e}^{i\alpha},\quad
\bra{\tilde{z}}=\frac{\bar{z}}{|z|}=\mbox{e}^{-i\alpha}
\ \Longrightarrow\
\ket{\tilde{z}}\bra{\tilde{z}}=\braket{\tilde{z}}=1,
\]
we have
\begin{eqnarray}
\mbox{e}^{X_{0}}\mbox{e}^{X_{2}}
&=&\left(
\begin{array}{cc}
\mbox{e}^{i\theta_{1}} &        \\
   & \mbox{e}^{i\theta_{2}}
\end{array}
\right)
\left(
\begin{array}{cc}
\mbox{cos}(|z|) & \mbox{e}^{i\alpha}\mbox{sin}(|z|)    \\
-\mbox{e}^{-i\alpha}\mbox{sin}(|z|) & \mbox{cos}(|z|)
\end{array}
\right)        \nonumber \\
&=&
\left(
\begin{array}{cc}
\mbox{e}^{i\theta_{1}} &        \\
   & \mbox{e}^{i\theta_{2}}
\end{array}
\right)
\left(
\begin{array}{cc}
\mbox{e}^{i\alpha/2} &        \\
   & \mbox{e}^{-i\alpha/2}
\end{array}
\right)
\left(
\begin{array}{cc}
\mbox{cos}(|z|)  & \mbox{sin}(|z|)    \\
-\mbox{sin}(|z|) & \mbox{cos}(|z|)
\end{array}
\right)
\left(
\begin{array}{cc}
\mbox{e}^{-i\alpha/2} &        \\
   & \mbox{e}^{i\alpha/2}
\end{array}
\right)    \nonumber \\
&=&
\left(
\begin{array}{cc}
\mbox{e}^{i(\theta_{1}+\alpha/2)} &        \\
   & \mbox{e}^{i(\theta_{2}-\alpha/2)}
\end{array}
\right)
\left(
\begin{array}{cc}
\mbox{cos}(|z|)  & \mbox{sin}(|z|)    \\
-\mbox{sin}(|z|) & \mbox{cos}(|z|)
\end{array}
\right)
\left(
\begin{array}{cc}
\mbox{e}^{-i\alpha/2} &        \\
   & \mbox{e}^{i\alpha/2}
\end{array}
\right).
\end{eqnarray}
We can also obtain the form given in \cite{CJ}, which is called the Euler 
angle parametrization.

\par \vspace{3mm}
We conclude this note by making a comment on an application to quantum 
computation. We are interested in Quantum Computation.
One of key points of quantum computation is to construct some unitary 
matrices (called quantum logic gates) in an efficient manner like Discrete 
Fourier Transform when $n$ is large enough. 
However, such a quick construction is not easy, see \cite{nine}, \cite{KF1}, 
or see \cite{KF2}, \cite{KuF}, \cite{Tilma} for a qudit case.
The parametrization of unitary matrices given by Jarlskog may be convenient 
for our real purpose.

We also make another comment that the parametrization method is similar to the 
idea in \cite{Lloyd}, namely the $(j-1)\times (j-1)$ block in 
(\ref{eq:fundamental}) can be written as
\begin{eqnarray*}
{\bf 1}_{j-1}-\left(1-\mbox{cos}(\sqrt{\braket{z_{j}}})\right)
\ket{\tilde{z}_{j}}\bra{\tilde{z}_{j}}
&=&
{\bf 1}_{j-1}-\ket{\tilde{z}_{j}}\bra{\tilde{z}_{j}}
+\mbox{cos}(\sqrt{\braket{z_{j}}})\ket{\tilde{z}_{j}}\bra{\tilde{z}_{j}}  \\
&=&
P_{j-1}(0)+\mbox{cos}(\theta_{j-1})P_{j-1}(1)
\end{eqnarray*}
with the notations in \cite{Lloyd}. Although 
$
P_{j-1}(0)+\mbox{cos}(\theta_{j-1})P_{j-1}(1)
$ 
is of course not 
$
P_{j-1}(0)+\mbox{e}^{-i\theta_{j-1}}P_{j-1}(1)
$, 
the method has a similar structure. Further study will be required. 

\vspace{5mm}
\noindent{\em Acknowledgment.}\\
The author wishes to thank Karol Zyczkowski for some useful comments and 
suggestions.


\end{document}